\documentclass[12pt]{iopart}
\usepackage{iopams}
\usepackage{graphicx}
\usepackage{hyperref}
\usepackage{color}

\begin{document}

\title{Non-equilibrium Landauer transport model for Hawking radiation from a black hole}

\author{P. D. Nation}
\address{Advanced Science Institute, RIKEN, Wako-shi, Saitama 351-0198, Japan}
\address{Department of Physics, University of Michigan, Ann Arbor, Michigan 48109-1040, USA}
\eads{pnation@riken.jp}
\author{M. P. Blencowe}
\address{Department of Physics and Astronomy, Dartmouth College, Hanover, New Hampshire 03755-3528, USA}
\author{Franco Nori}
\address{Advanced Science Institute, RIKEN, Wako-shi, Saitama 351-0198, Japan}
\address{Department of Physics, University of Michigan, Ann Arbor, Michigan 48109-1040, USA}
\date{\today}

\begin{abstract}
We propose that the Hawking radiation energy and entropy flow rates from a black hole can be viewed as a one-dimensional (1D), \textit{non-equilibrium} Landauer transport process.  Support for this viewpoint comes from previous calculations invoking conformal symmetry in the near-horizon region, which give radiation rates that are identical to those of a single 1D quantum channel connected to a thermal reservoir at the Hawking temperature.  The Landauer approach shows in a direct way the particle statistics independence of the energy and entropy fluxes of a black hole radiating into vacuum, as well as one near thermal equilibrium with its environment.  As an application of the Landauer approach, we show that Hawking radiation gives a net entropy production that is 50\% larger than that obtained assuming standard three-dimensional emission into vacuum.
\end{abstract}
\submitto{New Journal of Physics}
\maketitle
\section{Introduction}
One of the main achievements of quantum field theory in curved spacetime is the verification of the equilibrium thermodynamic description of black hole mechanics \cite{wald:1994}. In using this formalism \cite{hawking:1974,hawking:1975}, Hawking was able to provide a physical interpretation of the black hole temperature through the discovery of particle pair production at the horizon, while also solidifying the connection between black hole entropy and horizon surface area predicted a few years earlier \cite{bekenstein:1972}.  Subsequently, there has been a large body of work devoted to understanding this thermodynamic description of black holes and its deeper implications \cite{jacobson:1995, bousso:2002,susskind:2006}.  Yet at the same time, the \textit{non-equilibrium} thermodynamic properties of black holes, namely the steady-state flow, or transport, of energy and entropy via Hawking radiation, has received markedly less attention \cite{saida:2006,saida:2007}.

When viewed by an observer at spatial infinity, the metric of a non-rotating, uncharged black hole is given by the (1+3)-dimensional Schwarzschild metric.  Therefore, in the thermodynamic description of black holes, it is natural to assume that the emission of Hawking radiation corresponds to that of a three-dimensional (3D) thermal body obeying the Stefan-Boltzmann law.  However, recently there has been an increasing body of evidence suggesting that black hole emission is instead a 1D radiative process.  One indicator is the well-known near-horizon approximation under which the Schwarzschild metric of a black hole can be reduced to a (1+1)-dimensional Rindler space possessing infinite-dimensional conformal symmetry \cite{fabbri:2005}.  The ability to calculate the stress-energy tensor using conformal symmetry is the basis for standard derivations of the Hawking flux \cite{birrell:1982,brout:1995}.  More recently, it has been suggested that this conformal symmetry is responsible for the Hawking effect \cite{agullo:2010}, as it has been shown that this symmetry alone is sufficient to determine both the Hawking thermal spectrum \cite{jacobson:1993,iso:2007} and radiation flux \cite{robinson:2005}; the Hawking radiation is an inherently (1+1)-dimensional process.  This near-horizon conformal symmetry also reproduces the Bekenstein-Hawking form of the black hole entropy \cite{carlip:2002}, thus connecting to the other familiar dimensional reduction in black hole physics, namely the holographic principle \cite{bousso:2002}.

The first to focus on the entropic and information implications of a 1D evaporation process was Bekenstein \cite{bekenstein:2001}, who proposed that the entropy flow rate from a black hole is of the same form as that of a 1D quantum channel \cite{pendry:1983}, thus constraining the information flow from a black hole.  This same 1D channel description applies in the context of laboratory analogues of Hawking radiation \cite{schutzhold:2005,philbin:2008,nation:2009}, and it was noted that the power output from the analogue Hawking process coincides with the optimal energy current through a single quantum channel \cite{nation:2009}.  The concept of a 1D quantum channel was first considered by Landauer and others \cite{imry:1999,imry:2008} in the modeling of electrical transport in mesoscopic circuits.  The Landauer approach expresses the conductance of a 1D system $G_{c}$ in terms of its scattering properties \cite{imry:2008} via the relation
\begin{equation}\label{eq:landauer}
G_{c}=\frac{I}{\mu_{1}-\mu_{2}}=\frac{e^{2}}{\pi\hbar}T,
\end{equation}
where $I$ is the current through the 1D channel, $\mu_{1}$ and $\mu_{2}$ are the chemical potentials of the channel reservoirs, and $T$ is the transmission coefficient.  For perfect transmission, $T=1$, the channel conductance is given by $e^2/(\pi\hbar)$, a value that is independent of the microscopic, material nature of the channel, due to the mutual cancellation of the group velocity and density of states factors entering the current formula in 1D.  This Landauer formalism was subsequently extended to describe multiple channels \cite{buttiker:1985,sivan:1986}, as well as thermal transport \cite{sivan:1986,rego:1998,blencowe:1999,schwab:2000,meshke:2006}, where the currents are generated by temperature differences rather than by chemical potential differences.  Quantum mechanics places upper limits on the energy and entropy currents in 1D channels. These upper limits  are attained in the absence of backscattering for bosonic channels \cite{pendry:1983,blencowe:2000}, and are again independent of the material properties of the channel.  Furthermore, for thermal transport, these upper limits can be independent of whether the particles are bosons or fermions, and thus are termed ``universal" \cite{blencowe:2000,blencowe:2004}.  

Motivated by these connections, in this paper we argue that a non-equilibrium Landauer-transport model can be applied to black hole entropy flow and energy production rates, describing the Hawking effect in terms of currents flowing in 1D quantum channels connecting thermal reservoirs at each end. We thus relate the emission of Hawking radiation of astrophysical black holes to 1D thermal transport in mesoscopic devices; systems that differ by orders of magnitude in energy.  In particular, we emphasize the conditions under which the 1D currents are independent of particle statistics.  In contrast to the emitted power, the black hole entropy current cannot be obtained directly from the stress-energy tensor, and is rarely touched on in the literature without a priori assuming the validity of the 3D Stefan-Boltzmann law \cite{saida:2006,saida:2007,zurek:1982}.  Therefore, a theory that is capable of providing \textit{both} the black hole energy and entropy currents is required for the correct description of black hole evaporation \cite{thooft:1993}.

Assuming the validity of 1D Landauer transport theory enables the description of certain non-equilibrium, steady state emission processes for black holes, without necessarily requiring knowledge of their microscopic physics.  In essence, the Landauer approach allows us to extend the methodology of applying thermodynamic principles to black holes \cite{bekenstein:1974}.  Moreover, the Landauer model gives a physical insight into the transport of energy and entropy from a black hole that is lacking in existing field-theoretic derivations.

This paper is organized as follows.  In Sec.~\ref{sec:near} we review the well-known near-horizon approximation and the resulting conformal symmetry that leads to the standard derivation of the stress-energy tensor and the energy flow rate for Hawking radiation.  Next, in Sec.~\ref{sec:1d} we introduce the Landauer transport description for 1D quantum channels, and highlight the statistics-independent properties of the energy and entropy transport in these channels.  Section~\ref{sec:app} establishes the Landauer transport model to the emission of Hawking radiation, for both bosonic and fermionic particles, to a black hole in vacuum.  Charged and rotating black holes are also addressed.  As an application of the 1D Landauer approach, In Sec.~\ref{sec:entropy} we obtain the net entropy production of a black hole and compare with the standard 3D calculation given in Ref.~\cite{zurek:1982}.  The special case of a black hole near thermal equilibrium with its environment is also highlighted.  Finally, Sec.~\ref{sec:conclusion} ends with a brief discussion of the results.

\section{Near-horizon conformal symmetry and the Hawking flux}\label{sec:near}
For an observer near the horizon of a spherically symmetric Schwarzschild black hole of mass $M$, the original 4D metric $(G=c=1)$,
\begin{equation}\label{eq:schwar}
ds^{2}=-\left(1-\frac{2M}{r}\right)dt^{2}+\frac{dr^{2}}{\left(1-\frac{2M}{r}\right)}+r^{2}d\Omega^{2},
\end{equation}
can be reduced to that of a (1+1)-dimensional spacetime through the coordinate transformation $r=2M+x^{2}/8M$, where near $x=0$, $1-2M/r\approx x^{2}/16M^{2}$.  Near the horizon, excitations and dimensional quantities transverse to the $t$-$x$ plane are redshifted and can be ignored \cite{carlip:2007} (i.e. effective potentials for partial wave modes vanish exponentially fast at the horizon \cite{robinson:2005}).  Thus, the near-horizon form of the metric is given by \cite{fabbri:2005}
\begin{equation}\label{eq:rindler}
ds^{2}=-\left(\kappa x\right)^{2}dt^{2}+dx^{2},
\end{equation}
where $\kappa=1/4M$ is the surface gravity and the $t$-$x$ portion of the metric defines the flat (1+1)-dimensional Rindler spacetime.  Equation~(\ref{eq:rindler}) can be brought into conformal form by defining the coordinate $x=\kappa^{-1}\exp\left(\kappa \xi\right)$ and forming null coordinates, $u=t-\xi$ and $v=t+\xi$, under which the metric takes the form
\begin{equation}\label{eq:conformal}
ds^{2}=-C(u,v)du\,dv=-e^{\kappa\left(v-u\right)}du\,dv,
\end{equation}
where $C(u,v)$ is the conformal factor.  Here we ignore the effects of the radial potential as it is blue-shifted away by the conformal symmetry \cite{agullo:2010}.  The regularized expectation values for the stress-energy tensor can be immediately evaluated from the conformal structure of Eq.~(\ref{eq:conformal}): \cite{padmanabhan:2005}
\begin{equation}\label{eq:conformal-expect}
\left<T^{2\mathrm{D}}_{ii}\right>=-\frac{1}{12\pi}C^{1/2}\partial_{i}^{2}C^{-1/2},
\end{equation}
for $i=u,v$.  For a Schwarzschild black hole, the expectation value with respect to the Unruh vacuum at the horizon, for a single photon polarization, is given as \cite{brout:1995}
\begin{equation}\label{eq:influx}
\left.\left<T^{2\mathrm{D}}_{vv}\right>_{U}\right|_{r=2M}=-\frac{1}{12\pi}\left(\frac{1}{64M^{2}}\right)=-\frac{\pi}{12\hbar}T_{\mathrm{H}}^{2},
\end{equation}
where $T_{\mathrm{H}}=\kappa/2\pi$.  This represents the influx of negative energy across the horizon, responsible for the evaporation of the black hole, corresponding to the outgoing Hawking flux, as may be checked using the conformal factor for the t-r sector of the Schwarzschild metric, $C(r)=\left(1-M/r\right)$, and Eq.~(\ref{eq:conformal-expect})
\begin{equation}\label{eq:outflux}
\left<T^{2\mathrm{D}}_{uu}\right>_{U}=\frac{\pi}{12}T_{\mathrm{H}}^{2}\left[1-\frac{2M}{r}\right]^{2}\left[1+\frac{4M}{r}+\frac{12M^{2}}{r^{2}}\right] .
\end{equation}
The power emitted through Hawking radiation as seen by an inertial observer at $r=\infty$ is obtained from Eq.~(\ref{eq:outflux}) as 
\begin{equation}\label{eq:hflux}
\left<T^{2\mathrm{D}}_{uu}\right>_{U}=\frac{\pi k_{\mathrm{B}}^{2}}{12\hbar}T_{\mathrm{H}}^{2},  
\end{equation}
where, reintroducing dimensional constants for later convenience, we have $T_{\mathrm{H}}=\hbar c^{3}/8\pi k_{\mathrm{B}}GM$.  With $\sim 98\%$ of photons, and likewise $\sim 96\%$ of neutrinos, emitted in the radial direction (s-wave) \cite{page:1976}, Eq.~(\ref{eq:hflux}) is approximately valid in the full 4D spacetime as well, where the stress-energy tensor in the $r$-$t$ plane is given as \cite{brout:1995}
\begin{equation}
\left<T^{4\mathrm{D}}_{\mu\nu}\right>=\frac{1}{4\pi r^{2}}\left<T^{2\mathrm{D}}_{\mu\nu}\right>.
\end{equation}
The net flux across a spherical surface of radius $r$ is then given by $4\pi r^{2}\left<T^{4\mathrm{D}}_{\mu\nu}\right>$, which results in a net flux that is again expressed though Eq.~(\ref{eq:hflux}) \cite{padmanabhan:2005}.

\section{One-dimensional quantum channels}\label{sec:1d}
As a model for a single 1D quantum channel, we will consider two thermal reservoirs characterized by the temperatures $T_{\mathrm{L}}$ and $T_{\mathrm{R}}$ and with chemical potentials $\mu_{\mathrm{R}}$ and $\mu_{\mathrm{L}}$, respectively.  The reservoirs are coupled adiabatically through an effectively 1D connection supporting the bidirectional propagation of particles. The subscripts L and R denote the left and right thermal reservoirs respectively.  Here we will assume $T_{\mathrm{L}}>T_{\mathrm{R}}$ and that the transport through the 1D-connection is ballistic.

Although our focus is on fundamental fields/particles, for complete generality we will assume interpolating fractional statistics where the distribution function is \cite{wu:1994}
\begin{equation}
f_{g}\left(E\right)=\left[w\left(\frac{E-\mu}{k_{\mathrm{B}}T}\right)+g\right]^{-1},
\end{equation}
where $w(x)^{g}\left[1+w(x)\right]^{1-g}=\exp(x)$ with $x\equiv\left(E-\mu\right)/k_{\rm B}T$.  Here, $g=0$ and $g=1$ describe bosons and fermions respectively.  The individual  single-channel energy and entropy currents flowing from the left (L) and right (R) reservoirs may be written as \cite{blencowe:2000,blencowe:2004}
\begin{equation}\label{eq:edot}
\dot{E}_{\mathrm{L(R)}}=\frac{\left(k_{\mathrm{B}}T_{\mathrm{L(R)}}\right)^{2}}{2\pi\hbar}\int_{x_{\mathrm{L(R)}}^{0}}^{\infty}\!\!\!\!\!\!\!\!dx\left(x+\frac{\mu_{\mathrm{L(R)}}}{k_{\mathrm{B}}T_{\mathrm{L(R)}}}\right)f_{g}(x)
\end{equation}
and
\begin{eqnarray}\label{eq:sdot}
\dot{S}_{\mathrm{L(R)}}=&&-\frac{k_{\mathrm{B}}^{2}T_{\mathrm{L(R)}}}{2\pi\hbar}\int_{x_{\mathrm{L(R)}}^{0}}^{\infty}\!\!\!\!\!\!\!\!dx\left\{f_{g}\ln f_{g}+\left(1-gf_{g}\right)\ln(1-gf_{g})\right. \cr
&&\left.-\left[1+(1-g)f_{g}\right]\ln\left[1+(1-g)f_{g}\right]\right\},
\end{eqnarray}
where $x_{\mathrm{L(R)}}^{0}=-\mu_{\mathrm{L(R)}}/k_{\mathrm{B}}T_{\mathrm{L(R)}}$.  We define the zero of energy with respect to the longitudinal component of the kinetic energy.  For the case of bosons with $\mu_{\mathrm{L}}=\mu_{\mathrm{R}}=0$ (e.g. photons), the net power and entropy flow through the quantum channel, $\dot{E}^{\leftrightarrow}_{1D}=\dot{E}_{\mathrm{L}}-\dot{E}_{\mathrm{R}}$ and $\dot{S}^{\leftrightarrow}_{1D}=\dot{S}_{\mathrm{L}}-\dot{S}_{\mathrm{R}}$ respectively, become
\begin{equation}\label{eq:power}
\dot{E}^{\leftrightarrow}_{\mathrm{1D}}=\frac{\pi k_{\mathrm{B}}^{2}}{12\hbar}\left(T_{\mathrm{L}}^{2}-T_{\mathrm{R}}^{2}\right)
\end{equation}
and
\begin{equation}\label{eq:Sflow}
\dot{S}^{\leftrightarrow}_{\mathrm{1D}}=\frac{\pi k_{\mathrm{B}}^{2}}{6\hbar}\left(T_{\mathrm{L}}-T_{\mathrm{R}}\right).
\end{equation}
The emitted power Eq.~(\ref{eq:power}) holds for all bosonic quantum channels since the group velocity and density of states mutually cancel in 1D.  

The unidirectional power 
\begin{equation}\label{eq:unip}
\dot{E}^{\rightarrow}_{1\mathrm{D}}=\frac{\pi k_{\mathrm{B}}^{2}T^{2}_{\mathrm{L}}}{12\hbar}
\end{equation}
and the entropy current 
\begin{equation}\label{eq:uniS}
\dot{S}^{\rightarrow}_{\mathrm{1D}}=\frac{\pi k_{\mathrm{B}}^{2}}{6\hbar}T_{\mathrm{L}}
\end{equation}
are the maximum possible rates for single-channel bosonic flow.
The unidirectional entropy current (\ref{eq:uniS}) is in fact the maximum possible rate for single-channel fermionic flow as well, i.e., it is independent of the particle statistics \cite{anghel:2002,blencowe:2004}.  To see this, we make a change of integration variables in Eq.~(\ref{eq:sdot}), $x=\left(E-\mu\right)/k_{\mathrm{B}}T\rightarrow w$, upon which the entropy current can be simplified to \cite{blencowe:2004}
\begin{equation}\label{eq:sw}
\dot{S}_{\mathrm{L}}=\frac{k_{\mathrm{B}}^{2}T_{\mathrm{L}}}{2\pi\hbar}\int_{w_g\left(\frac{-\mu_{\mathrm{L}}}{k_{\mathrm{B}}T_{\mathrm{L}}}\right)}^{\infty}dw\left[\frac{\ln\!\left(1+w\right)}{w}-\frac{\ln\! w}{1+w}\right].
\end{equation}
We can see that the statistics of the particles shows up only in the lower integration bound of Eq.~(\ref{eq:sw}).  The maximum current (\ref{eq:uniS}) is obtained in the degenerate limit where the statistics-dependence vanishes, since $-\mu_{\mathrm{L}}/k_{\mathrm{B}}T_{\mathrm{L}}\rightarrow 0^{+}$, $w_{g=0}(0)=0$ for bosons, and $-\mu_{\mathrm{L}}/k_{\mathrm{B}}T_{\mathrm{L}}\rightarrow -\infty$, $w_{g=1}(-\infty)=0$ for fermions.  However, this same statistics independence in the degenerate limit  does not hold for the unidirectional power Eq.~(\ref{eq:edot}).  If one instead considers bidirectional current flow for fermions with $\mu_{\mathrm{R}}=\mu_{\mathrm{L}}$ and $T_{\mathrm{R}}=0$, then in the degenerate limit one recovers the same maximum rate (\ref{eq:unip}) as for bosons \cite{blencowe:2000}.   If the maximum energy and entropy current expressions, Eqs.~(\ref{eq:unip}) and (\ref{eq:uniS}) respectively, are combined by eliminating $T_{\mathrm{L}}$, then one obtains equality for the bound 
\begin{equation}\label{eq:bound}
\left(\dot{S}^{\rightarrow}_{\mathrm{1D}}\right)^{2}\le\frac{\pi k_{\mathrm{B}}^{2}}{3\hbar}\dot{E}^{\rightarrow}_{\mathrm{1D}},
\end{equation} 
which holds for 1D quantum channels with arbitrary reservoir temperatures, chemical potentials, and particle statistics \cite{pendry:1983,blencowe:2000}.  We note in passing that this bound is similar in form to the conjectured Bekenstein holographic bound \cite{bekenstein:1981}.  

\section{Hawking radiation from a black hole in vacuum}\label{sec:app}
\begin{figure}[t]
\begin{center}
\includegraphics[width=5.5in]{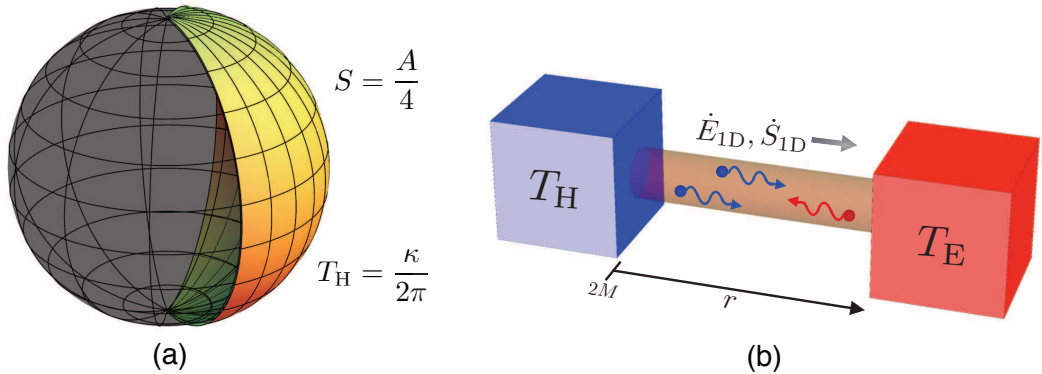}
\caption{(Color online) (a) In the equilibrium thermodynamic description of a Schwarzschild black hole, both the entropy $S$ and temperature $T_{\mathrm{H}}$ of the black hole are given by the properties of the two-dimensional horizon surface, a section of which is highlighted, being proportional to the surface area $A$ and surface gravity $\kappa$ respectively. (b) Near the horizon surface, the conformal symmetry results in an effectively (1+1)-dimensional spacetime, allowing for a 1D Landauer description.  Here, the power and entropy flow is through the 1D channel formed by the radial Schwarzschild coordinate $r$.  For a black hole in a thermal environment with temperature $T_{\mathrm{E}}>0$, the channel supports the bi-directional propagation of energy and entropy to and from the black hole. The net energy $\dot{E}^{\leftrightarrow}_{1\mathrm{D}}$ and entropy $\dot{S}^{\leftrightarrow}_{1\mathrm{D}}$ flow, Eqs.~(\ref{eq:power}) and (\ref{eq:Sflow}) respectively, is away from the black hole when $T_{\mathrm{H}}>T_{\mathrm{E}}$.}
\label{fig:fig1}
\end{center}
\end{figure}

The Landauer description of Hawking radiation is not limited to 1D, but also applies equally well to the 3D black hole spacetime viewed by an observer at infinity.  There, the entropy and energy flow rates can be characterized by a large ensemble of quantum channels, each labeled by a transverse spatial (i.e. angular momentum) quantum number, with interactions between channels described via a scattering matrix \cite{buttiker:1985}.  Therefore, scattering due to the potential barrier away from the horizon can be accounted for in the Landauer description through its known multichannel generalization with the inclusion of intra and inter-channel scattering (see, e.g., Ref.~\cite{sivan:1986}).  Although this seems to suggest that Hawking radiation flows through a vast number of quantum channels, the near horizon region, where Hawking radiation is emitted and absorbed, is not 3D but rather given by the Rindler metric, Eq.~(\ref{eq:rindler}).  With only a single spatial dimension remaining, the (1+1)-dimensional conformal symmetry of the metric near the horizon allows for a single 1D-quantum channel description of the Hawking process (see Fig.~\ref{fig:fig1}), where the remaining quantum channel corresponds to the lowest possible angular momentum mode.  Comparing Eq.~(\ref{eq:power}) with Eq.~(\ref{eq:hflux}), we can see immediately that the Landauer 1D channel formula for the zero chemical potential, bosonic power flow coincides with the Hawking radiation flux where $T_{\mathrm{L}}=T_{\mathrm{H}}$ and $T_{\mathrm{R}}=T_{\mathrm{E}}=0$, with $T_{\mathrm{E}}$ defined to be the temperature of the thermal environment surrounding the black hole.  The mutual cancellation of the group velocity and density of states factors in the 1D Landauer formula should make Eq.~(\ref{eq:unip}) valid not just in flat but in arbitrary curved spacetimes \cite{bekenstein:2001-2}, although the conformal symmetry of the near-horizon region suggests that the production of Hawking radiation is itself essentially a flat-space process.

Although we have appealed to conformal symmetry, these 1D emission properties of Hawking radiation are evident in the full 3D spacetime as well.  Following the original argument of Bekenstein \cite{bekenstein:2001} we note that the flat spacetime entropy emission rate for a blackbody in D-dimensions scales with the output power as 
\begin{equation}\label{eq:D}
\dot{S}_{\mathrm{D}}\propto \left(\dot{E}_{\mathrm{D}}\right)^{\mathrm{D}/(\mathrm{D}+1)}.  
\end{equation}
As a result, if a black hole were to radiate as a 3D object, one should expect the emitted entropy to scale as the $3/4$ power of the energy flow rate.  However, substitution of the Hawking temperature $T_{\mathrm{H}}$ into the Stefan-Boltzmann law, and making use of the black hole surface area $A=16\pi(GM)^{2}/c^{4}$, one finds that the emitted entropy 
\begin{equation}\label{eq:bhineq}
\dot{S}^{2}=\frac{1}{90}\frac{\pi k_{\rm B}^{2}}{3\hbar}\dot{E}
\end{equation}
goes as the $1/2$ power of the energy rate, just as one would expect for a 1D emitter.  In fact, Eq.~(\ref{eq:bhineq}) is identical to Eq.~(\ref{eq:bound}) up to a numerical factor arising from the assumption of a 3D, as opposed to 1D, emitter.  This result is attributable to the inverse dependence of the Hawking temperature on the black hole mass $M$, a property of black holes not shared by other blackbody emitters.  Therefore, the thermodynamic properties of a black hole correspond to that of a 1D blackbody emitter, as one might suspect given the ability to derive both the Hawking temperature and black hole entropy from the 1+1-dimensional conformal symmetry in the near horizon region.
  
In what follows, we will assume the validity of Eqs.~(\ref{eq:edot}) and (\ref{eq:sdot}) [equivalently Eq.~(\ref{eq:sw})] in describing the net energy and entropy outflow rates, respectively, for particles radiating from a black hole into the vacuum (i.e., $T_{\mathrm{E}}=0)$.  With the goal of introducing the Landauer description of Hawking radiation in the near horizon region, we will ignore scattering due to the radial potential barrier.  However, the full Landauer approach, relating transport to scattering processes \cite{imry:1999,imry:2008}, can incorporate inter-channel scattering due to particle interactions and back scatter from the radial potential barrier not considered here.  

The electrochemical potential of the black hole reservoir is $\mu_{\mathrm{L}}=\mu_{\mathrm{BH}}=q\Phi$, where $q$ is the electric charge of the field under consideration and $\Phi$ is the electrostatic potential corresponding to the charge of the black hole \cite{fabbri:2005}.  For a Schwarzschild black hole with $\Phi=0$, and hence $\mu_{\mathrm{BH}}=0$, bosons such as photons and gravitons have maximum rates given by Eq.~(\ref{eq:unip}) and (\ref{eq:uniS}) with $T_{\mathrm{L}}=T_{\mathrm{H}}$. For fermions such as neutrinos and electrons (i.e. leptons), setting $\mu_{\mathrm{BH}}=0$ gives a lower integration limit of $w_{g=1}(0)=1$ in Eq.~(\ref{eq:sw}), resulting in entropy and energy rates that are reduced by a factor of $1/2$ from the maximum values (\ref{eq:uniS}) and (\ref{eq:unip}).  This result for the energy rate was established in earlier calculations for massless fermions \cite{davies:1977}, and shows up in the relative values of the conformal and gravitational anomalies \cite{robinson:2005}.  However, as explicitly pointed out in Ref.~\cite{davies:1977}, the physical reason behind this result could not be established.  In contrast, the Landauer model presented here shows that these reduced fermionic currents are a direct consequence of the vanishing chemical potential of a Schwarzschild black hole and the 1D nature of the emission process.   Subsequently, it was pointed out \cite{davies:1978} that in a (1+1)-dimensional curved spacetime, the fermionic field describing a massless particle plus its antiparticle is equivalent to a single massless bosonic field.  From the Landauer viewpoint, the combined fermionic particle/antiparticle single channel currents can therefore be thought of as a single effective bosonic channel that satisfies the maximum rates, Eqs.~(\ref{eq:unip}) and (\ref{eq:uniS}), when $\mu_{\mathrm{BH}}=0$.  Although leptons are massive particles, the conformal symmetry removes the length scale set by the particle mass \cite{agullo:2010}; the particles are effectively massless.  In the case of ballistic transport, multiple channels can be treated independently.  Thus, the net  Schwarzschild black hole energy and entropy outflow rates are bounded by $N\left(T_{\mathrm{H}}\right)$ times the single channel rates given by Eq.~(\ref{eq:unip}) and Eq.~(\ref{eq:uniS}), respectively; a Schwarzschild black hole in vacuum radiates energy and entropy at the maximum rate allowed by quantum mechanics in 1D, saturating the bound in Eq.~(\ref{eq:bound}). Here, $N\left(T_{\mathrm{H}}\right)$ is the total number of effective bosonic channels spontaneously produced by a black hole at temperature $T_{\mathrm{H}}$; a quantity limited by the number of particle species emitted and their corresponding number of polarizations.  The temperature dependence of the effective channel number arises due to the requirement that $k_{\mathrm{B}}T_{\mathrm{H}}\gtrsim 2mc^{2}$ for pair production of particles with mass $m$.  

For a black hole with nonzero electrochemical potential, charged particle/antiparticle rates differ so as to cause the black hole net charge to decrease over time. The maximum entropy rate for a single charged fermionic channel coincides with the maximum rate for a single bosonic channel as shown above, giving Eq.~(\ref{eq:uniS}). The extent to which these maximum rates can be achieved depends on how close to degenerate is the thermal Hawking reservoir of the black hole for charged particles.  A special case is provided by extremal charged black holes \cite{fabbri:2005} satisfying $Q^{2}/M^{2}\approx1$, where $Q$ is the non-dimensional black hole charge.  In this limit, $T_{\mathrm{H}}\rightarrow 0$ giving $-\mu_{\mathrm{BH}}/k_{\mathrm{B}}T_{\mathrm{H}}\rightarrow -\infty$, the degenerate limit for fermions.  Charged fermions then satisfy Eq.~(\ref{eq:uniS}).  It may be possible to reach the degenerate limit for other choices of black hole parameters.  Similar reasoning applies to a black hole with angular momentum where, although spherical symmetry is broken, the emission of Hawking radiation is still governed by (1+1)-dimensional conformal symmetry \cite{agullo:2010}.  Here, the $U(1)$ gauge symmetry corresponding to the angular isometry in the (1+1)-dimensional theory may be written as a chemical potential in the same manner as that of a charged black hole \cite{iso:2006,iso:2006b}.  Therefore, the Landauer model presented here is quite general, being valid for black holes both with or without charge and angular momentum.  Finally, we point out that the cancellation of the density of states and group velocity in 1D quantum channels suggests that Eq.~(\ref{eq:unip}) should also be valid for black holes in other spacetimes, such as BTZ black holes \cite{banados:1992} in anti de-Sitter space, where conformal methods may still be applied \cite{carlip:2005}.

\section{Net entropy production in (1+1)-dimensions}\label{sec:entropy}
Originally considered by Zurek \cite{zurek:1982}, the rate of net entropy production by a Schwarzschild black hole due to the emission of Hawking radiation into a thermal environment, neglecting backscattering due to the radial potential barrier \cite{page:1983}, is given by  
\begin{equation}\label{eq:R}
R=\frac{dS}{dS_{\mathrm{BH}}}=T_{\mathrm{H}}\frac{dS}{dE}=T_{\mathrm{H}}\frac{\dot{S}}{\dot{E}}
\end{equation}
where we have used the first law of thermodynamics $dE_{\mathrm{BH}}=T_{\mathrm{H}}dS_{\mathrm{BH}}$ and assumed energy conservation, $dE=dE_{\mathrm{BH}}$.  For a 3D black hole radiating into a thermal environment with temperature $T_{\mathrm{E}}$, the power and entropy currents are 
\begin{eqnarray}
\dot{E}^{\leftrightarrow}_{\mathrm{3D}}&\sim& a\left(T_{\mathrm{H}}^{4}-T_{\mathrm{E}}^{4}\right)\\
\dot{S}^{\leftrightarrow}_{\mathrm{3D}}&\sim& \frac{4a}{3}\left(T_{\mathrm{H}}^{3}-T_{\mathrm{E}}^{3}\right),
\end{eqnarray}
where $a$ is a constant. Upon substitution into Eq.~(\ref{eq:R}), this yields the 3D black hole entropy production ratio
\begin{equation}\label{eq:3d}
R_{\mathrm{3D}}=\frac{4}{3}\frac{1-\left(T_{\mathrm{E}}/T_{\mathrm{H}}\right)^{3}}{1-\left(T_{\mathrm{E}}/T_{\mathrm{H}}\right)^{4}},
\end{equation}
which gives $R_{\mathrm{3D}}=4/3$ for a black hole in vacuum: $T_{\mathrm{E}}/T_{\mathrm{H}}\rightarrow 0$ \cite{zurek:1982}.  

However, as we have shown above, the emission properties of a black hole are better characterized as a 1D Landauer process.  Therefore, we compare the entropy produced via our 1D model to the standard calculation presented by Zurek.  Our focus in this paper is on the near-horizon region, and hence we do not include scattering.  The conformal symmetry in this region removes any inherent length scales, allowing the scattering barrier to be blue shifted away \cite{agullo:2010}.  Moreover, it is important to note that  the entropy current Eq.~(\ref{eq:sw}), like the 1D energy flow Eq.~(\ref{eq:unip}), should hold for a black hole in any spacetime where conformal symmetry may be invoked, even though the corresponding scattering properties may be markedly different.  Likewise, the Landauer approach is valid for analogue black hole models \cite{schutzhold:2005,philbin:2008,nation:2009}, as well as a moving mirror in (1+1)-dimensional spacetime \cite{fulling:1976} that can reproduce the emission properties of Hawking radiation from a Schwarzschild black hole in vacuum, even though no scattering barrier is present.  The effects of any scattering potential can be incorporated into a multichannel Landauer model \cite{buttiker:1985,sivan:1986}, and for the current case of a Schwarzschild black hole, will be presented elsewhere.  Since scattering can serve to only increase the net entropy produced from a Schwarzschild black hole \cite{page:1983}, the entropy production rates considered in this section may be viewed as lower bounds.  However, we note that for 1D transport, scattering will reduce the individual unidirectional energy $\dot{E}_{\mathrm{L(R)}}$ and entropy currents $\dot{S}_{\mathrm{L(R)}}$, Eq.~(\ref{eq:edot}) and Eq.~(\ref{eq:sdot}) respectively, to values below the ballistic bosonic channel limits, Eqs.~(\ref{eq:unip}) and (\ref{eq:uniS}).

\begin{figure}[htbp]
\begin{center}
\includegraphics[width=3.0in]{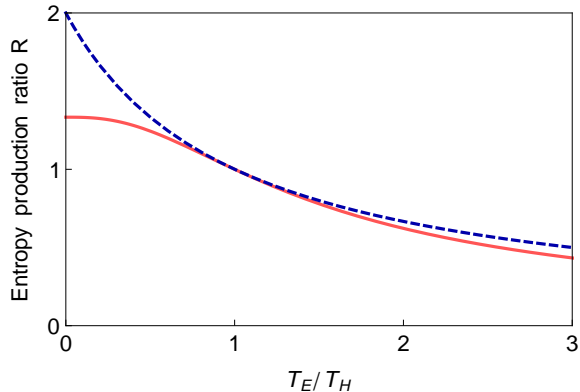}
\caption{(Color online) Entropy production ratio for a black hole characterized as 1D quantum channel $R_{\mathrm{1D}}$ (dashed-blue) compared to the standard 3D answer $R_{\mathrm{3D}}$ (red).  Both results agree near thermal equilibrium $T_{\mathrm{H}}\approx T_{\mathrm{E}}$. }
\label{fig:fig2}
\end{center}
\end{figure}

For comparison, in our Landauer model we set $\mu_{\mathrm{E}}=\mu_{\mathrm{BH}}=0$, and the net energy and entropy currents are given by Eqs.~(\ref{eq:power}) and (\ref{eq:Sflow}) respectively. The factors of 1/2 in the fermion rates will drop out when evaluating the ratio Eq.~(\ref{eq:R}).  The 1D entropy production ratio is then
\begin{equation}\label{eq:1d}
R_{\mathrm{1D}}=2 \frac{1-\left(T_{\mathrm{E}}/T_{\mathrm{H}}\right)^{\phantom{1}}}{1-\left(T_{\mathrm{E}}/T_{\mathrm{H}}\right)^{2}},
\end{equation}
which yields a larger value of $R_{1D}=2$ when radiating into vacuum; the net entropy production by Hawking radiation into vacuum is 50\% larger than that of a corresponding 3D thermal body at the Hawking temperature.  Again, this is due to the 1D properties of the near-horizon region, and the emitted radiation, for which Eq.~(\ref{eq:3d}) is no longer valid.  The difference between the 3D and 1D entropy rates, Eqs.~(\ref{eq:3d}) and (\ref{eq:1d}) respectively, for various ratios of $T_{\mathrm{E}}/T_{\mathrm{H}}$ is presented in Fig.~\ref{fig:fig2}.
In the case  where $T_{\mathrm{H}}\approx T_{\mathrm{E}}$, both Eqs.~(\ref{eq:3d}) and (\ref{eq:1d}) give approximately $R\approx 1+\delta/T_{\mathrm{H}}$ to first order in $\delta=\left(T_{\mathrm{H}}-T_{\mathrm{E}}\right)/2$.  As to be expected, in equilibrium ($\delta=0$), there is no net entropy production ($R=1$).

Near thermal equilibrium we can make use of linear response for small temperature differences, $\left(T_{\mathrm{H}}-T_{\mathrm{E}}\right)\ll\bar{T}$ where $\bar{T}=\left(T_{\mathrm{H}}+T_{\mathrm{E}}\right)/2$, to relate the energy and entropy flows by $\dot{S}_{1\mathrm{D}}=\dot{E}_{1\mathrm{D}}/\bar{T}$.  In this regime the unidirectional entropy rate Eq.~(\ref{eq:uniS})  allows us to recover the quantum of thermal conductance for a single effective bosonic channel \cite{blencowe:2004}:
\begin{equation}\label{eq:conductance}
G_{Q}=\frac{\dot{E}_{1\mathrm{D}}}{T_{\mathrm{H}}-T_{\mathrm{E}}}=\frac{\left(\dot{S}_{\mathrm{H}}-\dot{S}_{\mathrm{E}}\right)\bar{T}}{T_{\mathrm{H}}-T_{\mathrm{E}}}=\frac{\pi k_{\mathrm{B}}^{2}}{6\hbar}\bar{T},
\end{equation}
that, like Landauer's original expression Eq.~(\ref{eq:landauer}), relates conductance to transmission via only fundamental constants.  From the statistics independence of Eq.~(\ref{eq:uniS}), it follows that Eq.~(\ref{eq:conductance}) provides a general upper bound on the thermal conductance of a black hole that is independent of the particle statistics, as discussed in \cite{krive:1999,rego:1999}. 

\section{Conclusion}\label{sec:conclusion}
Using the conformal symmetry in the near-horizon region of a black hole, we have presented a 1D Landauer transport model for the non-equilibrium transport of both energy and entropy flow from a black hole, valid for particles with arbitrary statistics, and which clarifies the independence  of the underlying microscopic physics. Although our focus is on the near horizon region, the 1D nature of the emission properties are evident in the full (1+3)-dimensional spacetime seen by an observer at infinity, and may be derived from the inverse relationship between black hole mass and Hawking temperature. For a Schwarzschild black hole in vacuum, conformal symmetry results in a Hawking radiation energy flux that is identical to the power flowing in a single 1D quantum channel connected to a thermal bath with the Hawking temperature at one end and zero temperature at the other. Including multiple particle-species and polarizations, a Schwarzschild black hole in vacuum radiates power and entropy at the optimal rate, as a collection of effective bosonic channels. This is a direct result of the statistics independence of unidirectional energy and entropy flow in 1D highlighted by the Landauer formalism, and has not been discussed previously. Furthermore, we have shown that the reduced emission rates for fermions from a Schwarzschild black hole are due to the absence of a black hole chemical potential, giving a physical interpretation that is lacking in previous derivations. Moreover, in contrast to field-theory derivations using the stress-energy tensor, our Landauer model directly yields the entropy current from a black hole without assuming the validity of the 3D Stefan-Boltzmann law.  Both the charge and angular momentum of a black hole may be represented as an effective black hole chemical potential, and can be fully incorporated into the Landauer description presented here.  The unidirectional entropy current leads to a statistics independent heat flow near thermal equilibrium characterized by the quantum of thermal conductance.  Again, this property of black hole transport has not been addressed earlier.   In addition, the energy and entropy currents in 1D give a Hawking radiation entropy production ratio that is twice the corresponding value lost by the black hole when radiating into vacuum: a 50\% higher value when compared to the currently accepted 3D blackbody rate.  These results are a direct consequence of the reduced dimensionality in the near-horizon region and its conformal symmetry.  Given the intimate connection between entropy and information, the present findings, in particular Eq.~(\ref{eq:bound}), place strict limits on the rate of information transfer into and out of a black hole \cite{lloyd:2004}, and therefore will play a role in addressing the information loss problem in black hole evaporation \cite{hawking:1976,bekenstein:2004}.  However, we note that the reliance on conformal symmetry means that the Landauer model, in its present form, is incapable of describing non-thermal Hawking spectrum transport properties required for unitary black hole evaporation \cite{page:1993,nation:2010}. 

\ack
PDN thanks S. Carlip and H. Kang for helpful discussions.  PDN was supported by JSPS Postdoctoral Fellowship P11202. MPB was partially supported by the National Science Foundation (NSF) under grant No. DMR-0804477.  FN acknowledges partial support from DARPA, AFOSR, Laboratory of Physical Science, National Security Agency, Army Research Office, NSF grant No. 0726909, JSPS-RFBR contract No. 09-02-92114, Grant-in-Aid for Scientific Research (S), MEXT Kakenhi on Quantum Cybernetics, and JSPS through the FIRST program.

\bibliographystyle{unsrt}
\bibliography{text}
\end{document}